\documentstyle[preprint,aps]{revtex}
\begin{document}
\draft
\preprint{UCF-CM-95-002}
\title
{Integer quantum Hall effect for hard-core bosons and a failure of bosonic
Chern-Simons mean-field
theories for electrons at half-filled Landau level}
\author{O. Heinonen and M.D. Johnson}
\address{
Department of Physics, University of Central Florida, Orlando, FL 32816-2385
}
\maketitle
\begin{abstract}
Field-theoretical methods have been shown to be useful in constructing
simple effective theories for two-dimensional (2D) systems. These effective
theories are usually studied by perturbing around a mean-field approximation,
so the question whether such an approximation is meaningful arises immediately.
We here  study 2D interacting electrons in a half-filled
Landau level mapped onto interacting hard-core bosons in a magnetic field.
We argue that an interacting hard-core boson system in a uniform external
field such that there is one flux quantum per particle (unit filling)
exhibits an integer quantum Hall effect. As a consequence, the mean-field
approximation for mapping electrons at half-filling to a boson system
at integer filling fails.
(Contact mdj@physics.ucf.edu.  Cond-mat paper cond-mat/9510122.)
\end{abstract}
\pagebreak

Chern-Simons (CS) field theoretical approaches to the fractional
(and integer) quantum Hall effect (FQHE and IQHE, respectively) developed
recently\cite{Chern} provide simple effective theories for electron FQHE
and IQHE systems. In these approaches, the physical electron system
is mapped by a singular gauge transformation onto an equivalent system
of fermions or bosons interacting with a
statistical CS gauge field. The equivalent system is then typically studied by
first introducing a mean-field approximation, in which the particles experience
a constant net field which is the sum of the external field and the average
CS field, and then by perturbing around this mean-field approximation. Since
the CS field is singular, it may at first seem surprising that the rather crude
mean-field approximation makes sense at all, and it is essential to understand
when this approximation works, {\em i.e.\/} when perturbation about the
mean-field converges to the physical system, or at least gives physical
results, and when it breaks down. The key to success
of the mean-field approximation as a suitable starting point seems to be
when both the physical electron system and the equivalent CS system have
energy gaps. At the very least, perturbation about the mean-field then makes
sense, since the energy gap suppresses fluctuations about the mean-field,
although there is no guarantee that perturbation theory converges to the
physical system.
When both the physical system and the equivalent CS system are gapless,
sensible
results can be obtained from the mean-field approximation, although infrared
divergences may show up\cite{Halperin} and some renormalization scheme has to
be
devised. The purpose of this paper is add to our understanding of the
applicability of
CS theories by giving an example for which we argue that the mean-field
approximation is
not a suitable starting point at all: a half-filled Landau level mapped onto
CS interacting hard-core bosons at one filled Landau level.
For this case, we will argue that perturbation theory
about the mean-field does not converge to the physical system at all. This is
because the mean-field approximation describes a system of interacting
hard-core bosons at
integer filling. Jain and Rao\cite{JainRao} have recently suggested that
non-interacting bosons at $\nu=1$ have an energy gap and exhibit
an integer quantum Hall effect, and we argue that this holds for interacting
hard-core bosons. This
gap remains to all orders in perturbation theory, while the real electron
system is gapless.

Jain\cite{Jain} has
recently shown that mapping the electron system onto CS fermions
provides a natural way to study
the FQHE. In this mapping, one starts with a system of
two-dimensional (2D) electrons of density $n$ in an external field $A_\mu$
such that the filling factor $\nu=2\pi n\ell_B^2
=p/(2np+1)$, with $B_\mu=\epsilon_{\mu\nu\lambda}\partial_\nu A_\lambda$ and
the magnetic length $\ell_B^2=c/(eB)$,
and $p$ and $n$ integers. (We will use units in which $\hbar=c=1$ and
notation in which Greek indices denote time and space dimension, and Roman
indices denote space dimension with an implicit flat Minkowski metric
$g_{\mu\nu}$.
Summation over Greek indices will be implied unless stated otherwise.)
One then performs a singular
gauge transformation by attaching flux tubes with an even number $2n$ of flux
quanta to
each electron, which transforms the system into an equivalent one of
fermions in an external field $A_\mu$
plus the Chern-Simon field $a_\mu$ from the flux tubes. The resulting
equivalent system can then be studied by starting with a
mean-field approximation (the saddle-point approximation in a Lagrangian
formulation), in which external and statistical magnetic fields are
taken to be uniform at their spatial average. The mean-field system thus
consists of fermions in an average
magnetic field at {\em integer} filling $\nu=p$. Fluctuations in the
statistical field can then be added for example within a random-phase
approximation in
a Hamiltonian formulation, or the equivalent one-loop approximation in a
Lagrangian formulation by expanding the fields up to second order about the
saddle-point and then integrating out the fermion fields.
It is expected that perturbation theory, such as
the one-loop expansion, about the saddle point
in this case gives a good description of the
low-lying excitations of the real system (except
for the magnetoroton minimum)\cite{Zhang}. The reason is that the saddle-point
approximation
describes an IQHE system, which has an energy gap. This gap is the cyclotron
energy of the net
average magnetic field which has contributions from the average statistical
field and the external field. One can then formally integrate out the
high-energy modes to
obtain an effective theory of a massive fermion field coupled to a CS field.
It is well-known that the statistical
parameter (the topological mass term) of a Chern-Simons field
coupled to a massive scalar or spinor
field does not renormalize beyond the one-loop level, at which level it
receives
at the most a small correction\cite{renormalize}. Therefore, the energy gap
remains nonzero and finite to all orders in perturbation theory.
Furthermore, topological theorems\cite{Thouless} for the Hall conductivity
ensure that the statistical parameter does not renormalize {\em at all}
for IQHE and FQHE systems, so that the
incompressible system indeed has a quantized Hall conductivity. Note that it
is crucial that
the average magnetic field at the the saddle-point approximation
is not entirely due to the statistical field itself -- if this were the case,
the one-loop approximation would give rise to a {\em compressible} system due
to exact cancellation between the Hall conductivity of the CS fermions in the
constant average field and the statistical parameter of the CS term in the
Lagrangian.  This is precisely the case for anyon superconductors\cite{Zee},
where one starts with anyons in no external field and transform to CS
fermions or hard-core bosons in a net field due entirely to the CS field.

On the other hand, an FQHE system can be mapped onto
a system of hard-core CS bosons\cite{Zhang,Read}
by performing a singular
gauge transformation which attaches flux tubes with an odd number ($2p+1$) of
flux quanta to the electrons. At the mean-field level, one then
has a boson system
in a uniform magnetic field. In general, such a system is compressible
and contains vortices,
which are the locations where the boson order parameter vanishes, and these
generate an additional gauge field which has to be
included\cite{DHLee,Schmeltzer}. However,
at special fillings $\nu=1/(2p+1)$, {\em i.e.,} such that the average magnetic
field
precisely vanishes (the average Chern-Simon gauge field precisely cancels the
external gauge field), there are no vortices and the boson system is in
fact incompressible (see below). In this case,
the average vortex density vanishes\cite{DHLee}
and the vortices acquire a finite mass.
One can proceed to formally integrate out
the bose fields, and {\em assume}\cite{Zhang} that the
statistical parameter of the CS term does not renormalize.
The result
is an effective Lagrangian for the fluctuations in the CS field with only
massive modes and a quantized
Hall conductance. This non-trivial result has also been obtained
rigorously by Read\cite{Read} starting with the
Laughlin FQHE wavefunctions. The reason that the
statistical parameter does not renormalize (even though the
mean-field approximation
describes a system of hard-core bosons in zero magnetic field, which one
would think is gapless)
is that the FQHE states
have broken $U(1)$ symmetry described by off-diagonal long-range order in
a charged Bose field $\phi$ (the Bose field condenses into a superfluid).
The massless Goldstone
mode which appears in the broken-symmetry phase due to
phase fluctuations in the order parameter vanishes
by the standard Anderson-Higgs mechanism when the system is coupled to the CS
statistical field. Therefore, the effective CS Lagrangian has only massive
modes.

Recently, electron systems at half filling, $\nu=1/2$, have been studied
experimentally\cite{Wei,Engel,Koch} and theoretically.
Experimentally, these systems are gapless metals
and show a finite longitudinal resistivity $\rho_{xx}$ with a sharp minimum
as a function of filling factor precisely at $\nu=1/2$,
while the Hall resistivity $\rho_{xy}$ is {\em not}
quantized but rather shows its classical form $\rho_{xy}\sim B/(nec)$. Careful
experiments indicate\cite{Wei,Koch}
that in the presence of disorder, $\nu=1/2$ is the
critical point for a transition between an incompressible
quantum Hall fluid and an weakly localized Anderson insulator.
Theoretically, $\nu=1/2$ has been
studied by mapping the electrons onto CS fermions in zero average field
\cite{Kalmeyr,Halperin,Rezayi}, as well as by other more traditional
approaches\cite{others}.
Kalmeyr and Zhang\cite{Kalmeyr} showed that
within the CS theory, disorder in the effective system of 2D CS fermions in
zero average magnetic field leads to density fluctuations which couple to
fluctuations in the statistical field. This results in a fluctuating net
magnetic field which breaks the time-reversal invariance responsible for
weak localization in two dimensions, and the system shows a metallic
conductivity. In a very comprehensive and creative work, Halperin, Lee and
Read\cite{Halperin} investigated interacting CS fermions in zero average
magnetic field, with or without weak disorder. Under the assumption that
the system exhibits a Fermi surface at the mean-field level,
they went on to study the effective mass renormalization, in addition to
various
experimental signatures. Even though infrared divergences at the one-loop
level for short-range interactions
lead to infinite renormalization of the effective mass, while Coulomb
interactions lead to logarithmic corrections, they argued that the theory is
renormalizable and that the interacting system is Fermi-liquid like (or
is a marginal Fermi liquid). This
has recently been demonstrated in first-principle numerical calculations by
Rezayi and Read\cite{Rezayi}, and there is also recent experimental evidence
of a Fermi surface\cite{experiment1,experiment2,experiment3}.
It is clear that for this system, perturbation theory
about the mean-field system has to be done with extreme care -- at the
mean-field
level, the system is gapless (even though it is incompressible for Coulomb
interactions!\cite{Halperin}) and there is nothing which can guarantee
the stability of the one-loop
approximation and zero renormalization of the statistical parameter beyond the
one-loop level.
Therefore, one can expect heavy, perhaps infinite, renormalization of
all physical parameters, including the statistical parameter.

We will here study interacting electrons at $\nu=1/2$ in zero or weak
disorder by mapping the system onto a
system of CS hard-core bosons at filling $\nu=1$ by attaching a flux tube of
one flux quantum
to each electron. At the saddle-point, this is then a system of hard-core
bosons
at filling $\nu=1$. Jain and Rao\cite{JainRao} have recently argued that
non-interacting hard-core bosons exhibit a $\nu=1$ IQHE in the thermodynamic
limit. We will here argue that this holds when interactions are included.
As a consequence,
perturbation theory to infinite order about the
saddle point is renormalizable and converges, {\em but not to the physical
system}. We argue that this
is due to the non-analyticity of the true ground-state wavefunction, while the
mean-field wavefunction is analytical and outside the radius of convergence of
perturbation theory about the true ground state.

We start with a system of 2D electrons at half filling $\nu=1/2$ in the
external
field $A_\mu$. The system is described by the
action
\begin{eqnarray}
S&=&\int d^2r\,dt\,\left\{\psi^\dagger({\bf
r},t)\left[i\partial_0-eA_0-\mu\right]\psi({\bf r},t)
-\frac{1}{2m}\psi^\dagger({\bf r},t)
\left[-i\nabla-e{\bf A}\right]^2\psi({\bf r},t)
\right\}\nonumber\\
&&-\frac{1}{2}\int d^2r\,dt\int d^2r'\,rdt'\left(\psi^\dagger({\bf r},t)
\psi({\bf r},t)
-n_0\right)V(|{\bf r}-{\bf r}'|)\left(\psi^\dagger({\bf r}',t')\psi({\bf
r}',t')
-n_0\right),
\end{eqnarray}
where $n_0$ is the average density; $2\pi n_0\ell_B^2=\nu=1/2$.
By performing a singular gauge transformation
which attaches a flux quantum to each electron we obtain the Bose action
\begin{eqnarray}
S_B&=&\int d^2r\,dt\,\left\{\phi^\dagger({\bf
r},t)\left[\partial_0-eA_0-ea_0-\mu\right]\phi({\bf r},t)
-\frac{1}{2m}\phi^\dagger({\bf r},t)
\left[-i\nabla-e{\bf A}-e{\bf a}\right]^2\phi({\bf r},t)
\right\}\nonumber\\
&&+\frac{1}{2}\int d^2r\,dt\int d^2r'\,rdt'
\left(\phi^\dagger({\bf r},t)\phi({\bf r},t)
-n_0\right)V(|{\bf r}-{\bf r}'|)\left(\phi^\dagger({\bf r}',t')\phi({\bf
r}',t')
-n_0\right)\nonumber\\
&&+\frac{\theta e^2}{4}\epsilon_{\mu\nu\lambda}a_\mu f_{\nu\lambda}.
\end{eqnarray}
Here, the CS field $a_\mu$ is given by
\begin{equation}
{\bf a}({\bf r})=-\frac{c}{2\pi\theta e}\int d^2r'{\hat{\bf z}
\times({\bf r}-{\bf r'})\over
|{\bf r}-{\bf r}'|^2}
\phi^*({\bf r}')\phi({\bf r}'),
\end{equation}
with $\theta=1/(2\pi)$. We change to the Euclidean imaginary-time
action
\begin{eqnarray}
S_\tau&=&\int d^2r\,d\tau\,\left\{\phi^\dagger({\bf
r},\tau)\left[\partial_\tau-eA_0-ea_0-\mu\right]\phi({\bf r},\tau)
+\frac{1}{2m}\phi^\dagger({\bf r},\tau)
\left[-i\nabla-e{\bf A}-e{\bf a}\right]^2\phi({\bf r},\tau)
\right\}\nonumber\\
&&+\frac{1}{2}\int d^2r\,d\tau\int d^2r'\,r
\left(\phi^\dagger({\bf r},\tau)\phi({\bf r},\tau)
-n_0\right)V(|{\bf r}-{\bf r}'|)\left(\phi^\dagger({\bf r}',
\tau)\phi({\bf r}',\tau)
-n_0\right)\nonumber\\
&&+\frac{\theta e^2}{4}\epsilon_{\mu\nu\lambda}a_\mu f_{\nu\lambda}.
\end{eqnarray}
and consider the partition function
\begin{equation}
{\cal Z}=\int {\cal D}\phi^*\,{\cal D}\phi{\cal D}a_\mu\exp\left[-S_\tau\right]
\end{equation}
where the path-integrals are to be understood as coherent-states path
integrals.
We proceed by formally integrating out the bose fields with the CS fields
$a_\mu$ fixed
near their saddle-point values $a_\mu^0$, $\nabla \times{\bf a}^0=-2\pi n_0/e$,
to obtain the partition function
\begin{equation}
{\cal Z}_0(a_\mu)\int {\cal D}a_\mu\exp\left[-S_{\rm CS}(a_\mu)\right].
\end{equation}
Here,
$S_{\rm CS}(a_\mu)$ is the Euclidean Chern-Simons action. Next, we
expand the fields $a_\mu$ up to quadratic order about their saddle-point values
$a_\mu^0$. The result is an effective partition function ${\cal Z}_{\rm eff}
(a_\mu^0)$:
\begin{equation}
{\cal Z}_{\rm eff}(a_\mu^0)={\cal Z}_0(a_\mu^0)\int {\cal D}a_\mu'
\exp\left[-S_{\rm eff}(a_\mu')\right].\end{equation}
The effective action $S_{\rm eff}(a_\mu')$ describes the quadratic
fluctuations of
the CS field in a medium of hard-core bosons at a fixed magnetic field $\nu=1$
and is given formally by
\begin{equation}
S_{\rm eff}(a_\mu')=\int d^2r\,d\tau\,\frac{1}{2}[a_\mu']^*\Pi_{\mu\nu}a_\nu'
+\frac{\theta}{4}\epsilon_{\mu\nu\lambda}a_\mu'{f'}_{\nu\lambda},
\end{equation}
where $\Pi_{\mu\nu}$ is the current-current correlation function of the
interacting hard-core boson system at $\nu=1$,
\begin{equation}
\Pi_{\mu\nu}({\bf r},t;{\bf r}',t')=\langle j_\mu({\bf r},t)j_\nu({\bf r}',t')
\rangle.
\end{equation}

In general, $\Pi_{\mu\nu}$ is very complicated and describes properties such as
the dielectric function, magnetic susceptibility, and Hall conductivity
$\sigma_{xy}^0$,
and we don't know much at all about $\Pi_{\mu\nu}$. Except, as we now argue,
that this Bose system has an energy gap and that its
Hall conductivity is quantized at $\sigma_{xy}^0=e^2/(4\pi)$.
Jain and Rao\cite{JainRao} have recently suggested that non-interacting
hard-core bosons may exhibit IQHE at $\nu=1$ in
the thermodynamic limit. This may seem surprising, for the following reason.
Decompose the interaction potential $V(r-r')$ in relative angular momentum
(RAM) components,
the strength of which are given by the pseudo-potential parameters
$V_0,V_2,....$, where
$V_{2n}$ denotes the interaction energy of two bosons with RAM $2n$.
Note that the total wavefunction is even under interchange of two bosons, so
that no
two bosons can have odd RAM. The parameter $V_0$ describes the
hard-core interaction, since the zero RAM channel is the only one
which allows the bosons to be at the same place. We assume that
$V_0\gg\hbar\omega_c\gg V_2 >V_4>...$.
In the $V_0$ approximation, we can construct a Jastrow wavefunction in which
all bosons avoid zero RAM. With all bosons in the lowest Landau level, we must
have $\nu\leq1/2$ in
order to avoid RAM 0. By occupying the next Landau level, we can reach
fillings $\nu\leq2/3$ and
avoiding RAM 0, and so on. Simple trial wavefunctions corresponding to these
fillings can be
constructed by starting with the wavefunctions for non-interacting electrons
at $\nu=p$ and by
multiplying the wavefunction by the Jastrow factor $\prod(z_i-z_j)$. This
generates hard-core boson FQHE at
fillings $\nu=p/(p+1)$, with $\nu=1$ as an accumulation point, similar
to $\nu=1/2$ for
electrons. Consequently, one does not expect
a boson QHE at $\nu=1$. However, as $\nu\to1$ in this construction, there is
a finite fraction
of particles in all Landau levels, and the energy cost becomes very large.
Jain and Rao
suggested instead another variational scheme for boson wavefunctions, in which
the wavefunction is obtained from the product of
two fermion wavefunctions $\Psi_p$ and $\Psi_{p'}$,
each of which describes $p$ and $p'$ filled electron Landau levels,
respectively. For the case of non-interacting hard-core bosons at
$\nu=1$ the boson wavefunction $\Phi_1$
is then constructed from
two fermion wavefunctions for
filled two lowest Landau levels. The fermion wavefunction for two filled
Landau levels can be written
$\Psi_2=f_2([z,z*])\exp\left[-\sum_i|z_i|^2/4\right]$.
Here, $f_2([z,z*])$ is an antisymmetric polynomial
in the coordinates $z_i=x_i+iy_i$, and
$[z,z*]$ denotes the collections $\{z_i\}$ and $\{z_i^*\}$.
The resulting boson wavefunction
$\Phi_1=f_2^2\exp\left[-\sum_i|z_i|^2/4\right]$
at the very least gives a rigorous upper bound to
the ground state energy of bosons at $\nu=1$ which is much lower than
the previous one.
Although we have no formal proof, we will assume that $\Phi_1$ is
in fact the lowest-energy wavefunction at $\nu=1$ which avoids RAM 0.
This assumption can be supported by the following argument.
{}From the single-particle Landau level lowering operator
$\hat a\equiv1/\sqrt{2}\left(z/2+2\partial/\partial z^*\right)$
and angular momentum lowering operator
$\hat b\equiv1/\sqrt{2}\left(z^*/2+2\partial/\partial z\right)$
(we are using the symmetric gauge)
we construct many-body Landau level lowering and angular momentum lowering
operators $\hat A=\sum_i\hat a_i$ and $\hat B=\sum_i\hat b_i$ acting on the
many-body wavefunctions. A straightforward calculation then
yields $\hat A\Phi_1=\hat B\Phi_1=0$.
This is because both
operators act as derivative operators on $f_2^2$, which by construction is the
the square of the polynomial part of an electronic wavefunction of two filled
Landau levels. The net effect is then the same as having two Landau levels
filled with electrons
and attempting to two lower the total angular momentum or to send all electrons
to a lower Landau level, both of which give zero.
We conclude that $\Phi_1$
cannot be obtained by operating with any analytic function
of $\hat A^\dagger$ and $\hat B^\dagger$
on some other
wavefunction, and so is a good candidate for the ground state.

Within this variational scheme the state $\Phi_1$ has
an energy gap of order $\hbar\omega_c$ and so describes an IQHE. This
ground state and its energy gap
are robust under adiabatic turning-on of a finite number of RAM components
$V_{2n}$ since $\hbar\omega_c\gg V_{2n}$.
Using arguments due to
Laughlin \cite{Laughlin} and Halperin\cite{Halperin82}, it is
straightforward to demonstrate that this IQHE survives weak disorder, and that
$\sigma_{xy}^0$ is quantized at $\sigma_{xy}^0=e^2/(2\pi)$.
As a consequence of this, the effective statistical parameter $\theta_{\rm
eff}$
becomes\cite{Fradkin}
\begin{equation}
\frac{1}{\theta_{\rm eff}}=\frac{1}{\sigma_{xy}^0}+\frac{1}{\theta}
\end{equation}
or
\begin{equation}
\theta_{\rm eff}=\frac{e^2}{4\pi}.
\end{equation}

The energy gap at the saddle-point approximation ensures that the amplitude
of field
fluctuations about their saddle-point values are small, so a gradient
expansion of
$\Pi_{\mu\nu}$ makes sense to obtain the effective action for low-lying
excitations.
Since $\theta_{\rm eff}\not=0$, the low-lying modes are massive. In other
words, if
we were to start by intergrating out the short-distance fluctuations in the
Bose field, we would obtain
an effective theory describing a massive charged scalar field interacting
with the
CS field, and a non-zero statistical parameter $\theta$ which does not
renormalize to zero
at the one-loop level. Since $\theta$ does not renormalize at all beyond
the one-loop
level, this system will have an energy gap to all orders in perturbation
theory.
Consequently, perturbation theory about the saddle-point does not converge
to anything which describes the
original electron system, which is {\em gapless}.

This failure of perturbation theory about the saddle-point can be understood
as follows. The Hilbert space for the original electron wavefunction is
spanned by Slater
determinants of $N$ electrons in the $2N$ single-particle states in the
lowest Landau level (ignoring Landau-level mixing). The
electron wavefunction is thus a sum over $N$ products $(z_i-z_j)$ times
exponential factors. Thus, the wavefunction vanishes as $(z_i-z_j)$ as
two electrons
are brought towards one another. We then perform the singular gauge
transformation to
obtain the equivalent boson wavefunction. The gauge transformation only
changes the
relative phases of the wavefunction, which will now contain factors
$|z_i-z_j|$.
The singular gauge transformation maps the original analytic electron
wavefunction
in the lowest Landau level
onto a non-analytic boson wavefunction. Thus, the wavefunction so obtained does
not have an expansion in states in the lowest Landau level alone.
On the other hand,  the starting point for perturbation theory is
the homogeneous $\nu=1$ boson solution. The wavefunction of this state
$\Phi_1$ for
$\nu=1$ bosons in the large-$V_0$ limit is a product of factors $(z_i-z_j)$
and $z_i^*$,
times the exponential factors, which has a completely different non-analytic
structure. We therefore speculate
that the failure of perturbation theory about the saddle-point corresponds to
the fact
that the $V_0$ boson wavefunction at $\nu=1$ is outside the radius of
convergence
of perturbation expansions about the (exact) non-analytic boson wavefunction.

In conclusion, we have argued that interacting
hard-core bosons exhibit a $\nu=1$
IQHE, contrary to expectations. As a consequence, perturbation theory about the
saddle-point of the equivalent CS boson system at $\nu=1$ obtained by from
a singular gauge transformation of an electron system at $\nu=1/2$ fails
completely. The
reason is that the saddle-point approximation describes an incompressible
$\nu=1$ hard-core boson system with an energy gap.
When fluctuations at the one-loop level about the saddle point
are included, the statistical parameter remains non-zero, and it does not
renormalize
beyond the one-loop level. Therefore, the system has an energy gap and remains
incompressible to all orders in perturbation theory. This shows explicitly
that the
rather crude saddle-point approximation can give rise to completely
unphysical results.

We would like to thank the National Science Foundation for its support
through grant DMR93-01433, and O.H. would also like to thank
S. \"Ostlund and M. Jonson at Chalmers University of Technology for their
hospitality, where part of this work was done.

\end{document}